\documentclass[
twocolumn,
unsortedaddress,
superscriptaddress,
showpacs,
pra
]{revtex4}

\usepackage{amsmath}
\usepackage{amssymb}
\usepackage{bm}
\usepackage{amsthm}

\newtheorem{definition}{Definition}
\newtheorem{theorem}{Theorem}
\newtheorem{lemma}{Lemma}
\newtheorem{corollary}{Corollary}
\newtheorem{remark}{Remark}
\newtheorem{proposition}{Proposition}

\newcommand{\Def}{\stackrel{\textrm{def}}{=}}

\newcommand{\THEN}{\ensuremath{\Rightarrow}}

\newcommand{\IFF}{\ensuremath{\Leftrightarrow}} 

\newcommand{\C}{\ensuremath{{\mathcal{C}}}}

\newcommand{\E}{\ensuremath{{\mathcal{E}}}}
\newcommand{\F}{\ensuremath{{\mathbb{F}}}} 

\renewcommand{\H}{\ensuremath{{\mathcal{H}}}}
\newcommand{\I}{\ensuremath{{\mathcal{I}}}}
\newcommand{\J}{\ensuremath{{\mathcal{J}}}}
\newcommand{\K}{\ensuremath{{\mathcal{K}}}}
\renewcommand{\L}{\ensuremath{{\mathcal{L}}}}

\renewcommand{\O}{\ensuremath{{\mathcal{O}}}}
\renewcommand{\P}{\ensuremath{{\mathcal{P}}}}
\newcommand{\R}{\ensuremath{{\mathcal{R}}}}
\renewcommand{\S}{\ensuremath{{\mathcal{S}}}}

\newcommand{\Comp}{\ensuremath{{\mathbb{C}}}} 

\renewcommand{\epsilon}{\varepsilon}
\renewcommand{\phi}{\varphi}
\renewcommand{\d}{\mathrm{d}}
\newcommand{\etal}{\textit{et al.}~}
\newcommand{\wrt}{{w.r.t.~}}
\newcommand{\resp}{{resp.~}}

\DeclareMathOperator{\Tr}{Tr}
\newcommand{\TR}[1]{\mathrm{Tr}\bigl[#1\bigr]}
\newcommand{\PTR}[2]{\mathrm{Tr}_{#1}\bigl[#2\bigr]}
\newcommand{\EX}[2]{\mathrm{E}_{#1}\bigl[#2\bigr]}

\newcommand{\oneover}[1]{\frac{1}{#1}}
\newcommand{\defset}[2]{\left\{#1\left|\,#2\right.\right\}}

\newcommand{\cket}[1]{\left| #1 \right\rangle}
\newcommand{\bra}[1]{\left\langle #1 \right|}

\newcommand{\pure}[1]{%
\left| #1 \right\rangle\!\left\langle #1 \right|}
\newcommand{\inner}[2]{%
\left\langle #1, #2 \right\rangle}

\begin{document}

\title{
Quantum Secret Sharing Schemes and Reversibility of Quantum Operations
}

\author{Tomohiro Ogawa}
\email{ogawa@mist.i.u-tokyo.ac.jp}
\affiliation{
Graduate School of Information Science and Technology,
University of Tokyo,
7-3-1 Hongo, Bunkyo-ku, Tokyo, 113-8656 Japan.
}
\author{Akira Sasaki}
\email{Sasaki_Akira@dn.smbc.co.jp}
\affiliation{
Sumitomo Mitsui Banking Corporation,
3-2-1, Marunouchi, Chiyoda-ku, Tokyo 100-0004 Japan.
}
\author{Mitsugu Iwamoto}
\email{mitsugu@hn.is.uec.ac.jp}
\affiliation{
Graduate School of Information Systems,
University of Electro-Communications,
1-5-1 Chofugaoka, Chofu-shi, Tokyo, 182-8585 Japan.
}
\author{Hirosuke Yamamoto}
\email{Hirosuke@ieee.org}
\affiliation{
Department of Complexity Science and Engineering,
University of Tokyo,
P.Box 507, Transdisciplinary Sciences Bldg.,
5-1-5 Kashiwanoha, Kashiwa-shi, Chiba, 277-8561 Japan.
}

\date{\today}

\begin{abstract}
Quantum secret sharing schemes
encrypting a quantum state into a multipartite entangled state are treated.
The lower bound on the dimension of each share given by Gottesman
[Phys. Rev. A \textbf{61}, 042311 (2000)]
is revisited based on a relation between
the reversibility of quantum operations and the Holevo information.
We also propose a threshold ramp quantum secret sharing scheme
and evaluate its coding efficiency.
\end{abstract}



\maketitle

\section{Introduction}

\textit{Quantum secret sharing (QSS) schemes}
were studied by several authors
\cite{Hillery-et-al,Karlson-et-al,Cleve-Gottesman-Lo,Gottesman,Smith:QSS}
as quantum counterparts of
classical secret sharing (SS) schemes \cite{Shamir,Blakley}.
QSS schemes are methods
to encrypt an arbitrary quantum state or classical message
into a multipartite entangled state
among several quantum systems,
namely \textit{shares}, in the following way:
each of shares has no information about the original state or message
while it can be reproduced by collecting several shares.
QSS schemes can be classified into two categories
based on what is encrypted, i.e.,
quantum states \cite{Cleve-Gottesman-Lo,Gottesman,Smith:QSS}
or classical messages \cite{Hillery-et-al,Karlson-et-al}.
In this paper, we treat only the QSS schemes
encrypting quantum states,
which we call just QSS schemes for simplicity.

In the literature on QSS schemes in this sense,
the \textit{$(k,n)$-threshold QSS scheme} was proposed
by Cleve \etal \cite{Cleve-Gottesman-Lo}.
In the $(k,n)$-threshold QSS scheme,
an arbitrary quantum state is encoded into $n$ shares
so that any $k$ out of $n$ shares can reproduce the original state
while any $k-1$ or less shares have no information about it.
Recently, experimental demonstrations
\cite{Tittel-et-al:Experimental-demonstration-QSS,Lance-et-al:Tripartite-QSS}
of the threshold scheme were reported.
After the work of Cleve \etal,
Gottesman \cite{Gottesman} demonstrated that
any \textit{general access structure} consistent with
the \textit{monotonicity} \cite{Benaloh-Leichter}
and the no cloning theorem
\cite{Wootters-Zurek,Dieks,Yuen,Barnum-et-al,Koashi-Imoto:No-cloning}
can be realized by a QSS scheme.
The same result was shown by Smith \cite{Smith:QSS} independently
by using monotone span programs.
Gottesman \cite{Gottesman} also analyzed the coding efficiency
of QSS schemes and showed that the dimension of each share
must be the same or larger than that of the original system.

In this paper,
we revisit the coding efficiency of QSS schemes
in an information theoretical manner.
First, we establish a relation between
the \textit{reversibility} of quantum operations
and the Holevo information \cite{Holevo73}
in a general setting rather than QSS schemes.
This relation is a natural extension
of the idea in the classical information theory
that the \textit{sufficient statistic} is
characterized by the preservation of the mutual information
\cite{Cover-Thomas}.
In classical statistical inferences,
the sufficient statistic has
several equivalent characterizations \cite{Strasser}:
the existence of reverse channels,
the preservation of information quantities such as
the relative entropy, and the factorization theorem.
On the other hand,
the reversibility of quantum operations was studied
by several authors
\cite{Knill-Laflamme,Bennett-et-al,Schumacher-Nielsen}
related to the quantum error correcting code
\cite{Shor-pra52:QECC,Steane-prl77:QECC},
while
it was also studied in terms of \textit{sufficiency}
in the field of the operator algebra
\cite{Umegaki-III,Umegaki-IV,Gudder-Marchand,Hiai-Ohya-Tsukada-1981,Hiai-Ohya-Tsukada-1983,Petz1986:SufficientSubalgebra,Petz1988:SufficientChannels}
(see also
Refs.~\cite{Petz2003:MonotonicityRevisited,Hayden-Jozsa-Petz-Winter}).

Recently, Petz and his colleagues
\cite{Mozonyi-Petz-2003,Jencova-Petz-2004}
have established
a theory of sufficiency in the quantum setting,
that is characterized by the reversibility of quantum operations
(or coarse-grainings),
the preservation of information quantities,
and the quantum version
\cite{Koashi-Imoto:Operations,Lindblad:No-cloning}
of the factorization theorem.
Our characterization of the reversibility
falls into a natural variant of theirs.
However, we rather use the term reversibility in this paper
for the reasons that
the characterization is closely related to
the literature in the quantum error correcting code
and that the notion of sufficiency
is not yet so clear in quantum statistical inferences
such as the quantum estimation theory \cite{Helstrom:Text,Holevo:Text,AsymTheoQuantStat}
and quantum hypothesis testing
\cite{Holevo:StatisticalDecision,Helstrom:Text,Hiai-Petz:Stein,Ogawa-Nagaoka:Stein,AsymTheoQuantStat}.

Second, returning to QSS schemes,
we utilize the characterization of the reversibility
to evaluate a kind of information
that each share has about the original quantum state,
and then
the evaluation leads to the lower bound on the dimension of each share
given by Gottesman \cite{Gottesman}.
It should be noted
\footnote{
After the prototype \cite{SITA2003} of this work was concluded,
we were informed by an author of \cite{Anderson}
that they had established
a similar result on the dimension of each share.
}
that a similar result on the dimension of each share
has been given in Ref.~\cite{Anderson}
by a different method
using the reference system relevant to
the coherent information \cite{Schumacher-Nielsen}.

As mentioned above, it is impossible
to reduce the dimension of each share than that of the original system
in QSS schemes which have perfect security conditions.
Here, the perfect security conditions mean that
any set of shares can either reproduce the original state
or obtain no information about it,
and such schemes are called \textit{perfect QSS schemes}.
On the other hand, in classical ramp SS schemes
\cite{Yamamoto,Blakley-Meadows},
the size of each share can be decreased
by the sacrifice of security conditions
admitting the intermediate property for some sets of shares.
Following these classical counterparts,
we propose a \textit{ramp QSS scheme}
and analyze the coding efficiency of it.
Then, it is shown that the dimension of each share
can be reduced than that of the original system
by the sacrifice of security conditions
like the classical ramp schemes.
Finally, we also demonstrate an optimal construction of
the ramp QSS scheme.

\section{Definitions}

Let $\H,\J,\K$ be finite dimensional Hilbert spaces,
and let $\L(\H)$ and $\S(\H)$ be the totalities of
linear operators and density operators
on a Hilbert space $\H$, respectively.
We will treat QSS schemes encrypting a quantum state on $\H$
into a composite system of Hilbert spaces $\H_1,\dots,\H_n$,
each of which is called a \textit{share}.
Let $N\Def\{1,\dots,n\}$ be the entire set of shares
and $\H_N\Def\bigotimes_{i\in N}\H_i$ be the corresponding Hilbert space.
For a subset $X\subseteq N$ of shares,
let $\H_X\Def\bigotimes_{i\in X}\H_i$ as well.
The encoding operation of a QSS scheme is described by
a quantum operation $W_N: \S(\H)\rightarrow \S(\H_N)$,
which is a completely positive and trace preserving map.
For a subset $X\subseteq N$,
the composition map of the encoder $W_N$ and
the partial trace of the complement $N\backslash X$
is denoted by $W_X\Def\Tr_{N\backslash X}\cdot W_N$.

Now we will define the notion of the reversibility
for general quantum operations.
A quantum operation $\E:\S(\J)\rightarrow\S(\K)$ is called
\textit{reversible} with respect to (w.r.t.)
a subset $\S\subseteq \S(\J)$ of density operators
if there exists a quantum operation $\R:\S(\K)\rightarrow\S(\J)$
such that $\forall\rho\in\S,\,\R\cdot\E(\rho)=\rho$.
A quantum operation $\E:\S(\J)\rightarrow\S(\K)$
is called \textit{vanishing} \wrt $\S\subseteq \S(\J)$
if there exists a density operator $\rho_0\in\S(\K)$
such that $\forall\rho\in\S,\,\E(\rho)=\rho_0$.
\begin{remark}
\label{remark:1}
It should be noted here
that a quantum operation is reversible (\resp vanishing)
\wrt $\S\subseteq \S(\J)$
iff it is reversible (\resp vanishing) \wrt the extreme points
of the convex hull of $\S$.
Therefore,
letting $\S_1(\J)$ be the totality of pure states on $\J$,
a quantum operation is reversible (\resp vanishing) \wrt $\S(\J)$
iff it is reversible (\resp vanishing) \wrt $\S_1(\J)$.
\end{remark}

A QSS scheme is defined by
a quantum operation $W_N:\S(\H)\rightarrow \S(\H_N)$
which is reversible \wrt $\S(\H)$.
For a QSS scheme $W_N$,
a set $X\subseteq N$ is called
\textit{qualified} (\resp \textit{forbidden})
if $W_X$ is reversible (\resp vanishing) \wrt $\S(\H)$,
and, in addition,
a set $X\subseteq N$ is called \textit{intermediate}
if $W_X$ is neither reversible nor vanishing \wrt $\S(\H)$.
A QSS scheme $W_N$ is called a \textit{perfect scheme}
if any set $X\subseteq N$ is either qualified or forbidden.
Otherwise, $W_N$ is called a \textit{ramp scheme}.
Although the terms ``authorized'' and ``unauthorized'' are
used in the previous works \cite{Cleve-Gottesman-Lo,Gottesman}
on perfect QSS schemes,
we use the terms ``qualified'' and ``forbidden'' in this
paper because we must divide ``unauthorized'' sets between
``intermediate'' sets and ``forbidden'' sets in ramp QSS schemes.

\section{Access Structure}

The access structure of a QSS scheme is
the list of forbidden, intermediate, and qualified sets.
In classical ramp secret sharing (SS) schemes
\cite{Yamamoto,Blakley-Meadows},
intermediate sets are classified further into
multilevel categories based on the conditional entropy.
In ramp QSS schemes, however,
we do not classify the intermediate sets
for simplicity in this paper.
Formally, the access structure of the set $N$
is defined by a map $f: 2^N\rightarrow \{0,1,2\}$,
where $0$, $1$, and $2$ indicate forbidden, intermediate,
and qualified sets, respectively.
For a QSS scheme $W_N$,
the access structure of $N$ is determined naturally,
and hence, is called the access structure of $W_N$.
It is clear that the access structure of $W_N$ satisfies
the \textit{monotonicity}, i.e.,
$
X\subseteq Y \THEN f(X)\le f(Y).
$
In addition to this relation, the restriction due to
the no cloning theorem
\cite{Wootters-Zurek,Dieks,Yuen,Barnum-et-al,Koashi-Imoto:No-cloning}
(see also Proposition \ref{th:no-cloning-deleting} in the appendix)
is imposed on QSS schemes,
that is, the complement of a qualified set is necessarily forbidden.
Conversely, it was shown in Refs.~\cite{Gottesman,Smith:QSS}
that any perfect access structure, consistent with the monotonicity and
the no cloning theorem, can be realized by a perfect QSS scheme.

A quantum operation $\E$ is called a \textit{pure state channel}
if $\E(\rho)$ is a pure state for any pure state $\rho$.
A QSS scheme $W_N$ is called a \textit{pure state scheme}
if it is a pure state channel.
Otherwise, it is called a \textit{mixed state scheme}.
Gottesman \cite{Gottesman} showed that
any perfect QSS scheme is
regarded as a subsystem of a pure state QSS scheme.
The following lemma is a slight extension of his result
including ramp QSS schemes.

\begin{lemma}\label{lem:access-structure}
Any mixed state QSS scheme $W_N$ is realized
by discarding one share from a pure state QSS scheme $W_{N'}$.
Moreover, the access structure of $W_{N'}$ is determined uniquely
by that of $W_N$.
\end{lemma}
\begin{proof}
From the Stinespring dilation theorem \cite{Stinespring},
there exists a Hilbert space $\H_Z$ and an isometry
$V:\H\mapsto\H_N\otimes\H_Z$ such that
\begin{align}
W_N(\rho)=\PTR{Z}{V\rho V^*}.
\end{align}
Let $N'=N\cup Z$,
then $W_N$ is realized
from the pure state QSS scheme $W_{N'}(\rho)=V\rho V^*$
by discarding one share $Z$.
We note that the access structure of $W_{N'}$ for a set $X\subseteq N$
not including $Z$ is the same as that of $W_N$.
Hence we will consider the access structure of $W_{N'}$
for $X\subseteq N'$ which includes $Z$.
It follows from Proposition \ref{th:no-cloning-deleting} in the appendix
that $X$ is qualified iff $N'\backslash X$ is forbidden,
and equivalently that
$X$ is forbidden iff $N'\backslash X$ is qualified.
Furthermore, we can also see that
$X$ is intermediate iff $N'\backslash X$ is intermediate.
Therefore, the access structure $f(X)$ is determined uniquely by
the complement $N'\backslash X \subseteq N$.
\end{proof}

\section{Reversibility and Holevo Information}

In this section, turning to a general setting,
we will demonstrate that the Holevo information \cite{Holevo73}
is closely related to the reversibility of quantum operations.

For $\rho,\sigma\in\S(\J)$,
let
\begin{align}
D(\rho||\sigma)\Def\TR{\rho(\log\rho-\log\sigma)}
\end{align}
be the quantum relative entropy.
Then, for any quantum operation $\E:\S(\J)\rightarrow\S(\K)$,
it yields the monotonicity
\cite{Lindblad:CP,Araki:RelativeEntropy,Uhlmann:CP},
i.e.,
\begin{align}
D(\rho||\sigma)\ge D(\E(\rho)||\E(\sigma)),
\end{align}
and the equality holds iff
$\E$ is reversible \wrt $\{\rho,\sigma\}$
\cite{Petz1988:SufficientChannels}
(see also
Refs.~\cite{Petz2003:MonotonicityRevisited,Hayden-Jozsa-Petz-Winter}).
Furthermore, in the case of equality, there is
a canonical reverse operation depending only on $\sigma$,
which is given by
\begin{align}
\R_\sigma(\tau)
\Def\sigma^{\frac{1}{2}}
\E^*(\E(\sigma)^{-\frac{1}{2}}\tau\E(\sigma)^{-\frac{1}{2}})
\sigma^{\frac{1}{2}}
\label{canonical-reverse}
\end{align}
Here $\E^*:\L(\K)\rightarrow\L(\J)$ is the dual of $\E$ satisfying
\begin{align*}
\forall \rho\in\S(\J),\,\forall Y\in\L(\K),\,\TR{\E(\rho)Y}=\TR{\rho\E^*(Y)}.
\end{align*}
The above fact is summarized as the following proposition.
\begin{proposition}[Petz \cite{Petz1988:SufficientChannels}, see also Refs.~\cite{Petz2003:MonotonicityRevisited,Hayden-Jozsa-Petz-Winter}]
\label{th:divergence-equality}
Given a quantum operation $\E:\S(\J)\rightarrow S(\K)$
and $\rho,\sigma\in\S(\J)$,
let $\R_\sigma$ be the quantum operation defined by \eqref{canonical-reverse}.
Then the following three conditions are equivalent.
\begin{enumerate}
\item[(a)] $D(\rho||\sigma) = D(\E(\rho)||\E(\sigma))$
\item[(b)] $\R_{\sigma}\cdot\E(\rho)=\rho$
\item[(c)] $\E$ is reversible \wrt $\{\rho,\sigma\}$.
\end{enumerate}
\end{proposition}

This fact can be easily extended to a general relation between
the Holevo information and
the reversibility of a quantum operation
\wrt a subset $\S\subseteq\S(\J)$.
Let $\P(\S)$ be the set of probability measures on $\S\subseteq\S(\J)$,
and let
\begin{align}
\EX{\mu}{\,\cdot\,}=\int_{\S}\,\cdot\,\,\mu(\d\rho).
\end{align}
be the expectation by a probability measure $\mu\in\P(\S)$.
Given an ensemble $\mu\in\P(\S)$ and a quantum operation $\E$,
the Holevo information is defined by
\begin{align}
I(\mu;\E)
&\Def \EX{\mu}{D(\E(\rho)||\E(\sigma_{\mu}))}
\nonumber\\
&=H(\E(\sigma_{\mu}))-\EX{\mu}{H(\E(\rho))},
\end{align}
where $\sigma_{\mu}\Def \EX{\mu}{\rho}$
and $H(\rho)\Def-\TR{\rho\log\rho}$ is the von Neumann entropy.
Moreover,
let $\P_+(\S)$ be the set of probability measures on $\S\subseteq\S(\J)$
which are positive almost everywhere on $\S$. More specifically,
\begin{align}
\P_+(\S)
\Def
\defset{\mu\in\P(\S)}{\forall O\subseteq\O(\S),\,\mu(O)>0},
\end{align}
where $\O(\S)$ is the totality of open sets on $\S$,
and is defined in terms of
the relative topology induced by the inclusion $\S\subseteq\S(\J)$.
Then we have the following theorem.
\begin{theorem}\label{th:Holevo-info}
Let $\I:\S(\J)\rightarrow\S(\J)$ be the identity map.
Given a quantum operation $\E:\S(\J)\rightarrow\S(\K)$
and $\S\subseteq\S(\J)$,
the following three conditions are equivalent.
\begin{enumerate}
\item[(a)] $\E$ is reversible (\resp vanishing) \wrt $\S$.
\item[(b)] $\forall\mu\in\P_+(\S),\,I(\mu;\E)=I(\mu;\I)\;(\text{\resp} =0)$.
\item[(c)] $\exists\mu\in\P_+(\S),\,I(\mu;\E)=I(\mu;\I)\;(\text{\resp} =0)$.
\end{enumerate}
\end{theorem}
\begin{proof}
(a) \THEN (b):
From the definition of the reversibility,
there exists a quantum operation $\R$ such that
$\forall\rho\in\S,\,\R\cdot\E(\rho)=\rho$.
Taking the expectation of $\rho$ by an arbitrary $\mu\in\P_+(\S)$,
we have $\R\cdot\E(\sigma_{\mu})=\sigma_{\mu}$.
Then it follows from ``(c) \THEN (a)''
of Proposition \ref{th:divergence-equality} that
\begin{align}
\forall\rho\in\S,\,
D(\rho||\sigma_{\mu})=D(\E(\rho)||\E(\sigma_{\mu}))
\label{Holevo-1}
\end{align}
Taking the expectation of the above equality by $\mu$ leads to (b).

\noindent
(b) \THEN (c): Obvious.

\noindent
(c) \THEN (a):
First, note that
\begin{align}
I(\mu;\I)-I(\mu;\E)
&=\EX{\mu}{D(\rho||\sigma_{\mu})-D(\E(\rho)||\E(\sigma_{\mu}))}
\nonumber\\
&\ge 0,
\label{Holevo-2}
\end{align}
since the monotonicity of the quantum relative entropy leads to
\begin{align}
D(\rho||\sigma_{\mu})-D(\E(\rho)||\E(\sigma_{\mu})) \ge 0
\end{align}
for each term in the expectation of \eqref{Holevo-2}.
Therefore we can see from the definition of $\P_+(\S)$,
along with the continuity of $\E$ and the quantum relative entropy,
that
\eqref{Holevo-1} is a necessary condition for $I(\mu;\I)-I(\mu;\E)=0$.
Using ``(a) \THEN (b)'' of Proposition \ref{th:divergence-equality},
we have that
$\forall\rho\in\S,\,\R_{\sigma_{\mu}}\cdot\E(\rho)=\rho$,
which implies (a).

As for the vanishing property,
we can show the assertion in the same way as the reversibility
by using $D(\rho||\sigma)\ge 0$ 
and $D(\rho||\sigma)=0 \IFF \rho=\sigma$.
\end{proof}

\section{Coding Efficiency of QSS Schemes}

Let $\S_1(\H)$ be the totality of pure states on $\H$,
and note that a quantum operation is
reversible (\resp vanishing) \wrt $\S(\H)$
iff it is reversible (\resp vanishing) \wrt $\S_1(\H)$.
Therefore
it suffices to treat the reversibility
of a QSS scheme $W_N$ \wrt $\S_1(\H)$.
For a pure state ensemble $\mu\in\P_+(\S_1(\H))$,
the Holevo information is given by
$I(\mu;\I)=H(\sigma_{\mu})$,
since $H(\rho)=0$ for any pure state $\rho\in\S_1(\H)$,
and hence, the following theorem immediately follows from
Theorem~\ref{th:Holevo-info}.
\begin{theorem}\label{th:QSSS-info}
For any QSS scheme $W_N$, the following three conditions are equivalent.
\begin{enumerate}
\item[(a)] $X$ is qualified (\resp forbidden).
\item[(b)] $\forall\mu\in\P_+(\S_1(\H)),\,I(\mu;W_X)=H(\sigma_{\mu})\;
      (\text{\resp} =0)$.
\item[(c)] $\exists\mu\in\P_+(\S_1(\H)),\,I(\mu;W_X)=H(\sigma_{\mu})\;
      (\text{\resp} =0)$.
\end{enumerate}
\end{theorem}
\begin{remark}
\label{remark:2}
Theorem \ref{th:QSSS-info} can be regarded as
a variant of the perfect error correcting condition
\cite{Schumacher-Nielsen}
without using reference systems,
while Theorem \ref{th:Holevo-info} is an extension
of these conditions to
the reversibility condition \wrt general subsets of $\S(\H)$.
\end{remark}
\begin{remark}
\label{remark:3}
As is clear by definition,
it is also interesting to observe from Theorem \ref{th:QSSS-info}
that the access structure of QSS schemes does not depend
on $\mu$ in $\P_+(\S_1(\H))$.
We note that
this fact holds in classical perfect SS schemes.
Actually, corresponding statements are given
by a different approach in Ref.~\cite{Blundo-et-al}.
\end{remark}

Now we consider the coding efficiency of QSS schemes.
A set $X\subseteq N$ is called \textit{significant}
if there exists a forbidden set $Y\subseteq N$
such that $X\cup Y$ is qualified.
\begin{theorem}\label{th:QSSS-entropy-eval}
For any significant set $X\subseteq N$ of any QSS scheme $W_N$,
it holds that
\begin{align}
\forall\mu\in\P_+(\S_1(\H)),\,H(\sigma_{\mu})\le H(W_X(\sigma_{\mu})).
\end{align}
\end{theorem}
\begin{proof}
From Lemma \ref{lem:access-structure},
$W_N$ is supposed to be a pure state scheme
without loss of generality.
Moreover, for any significant set $X\subseteq N$
we can choose a forbidden sets $Y\subseteq N$
such that $X\cup Y$ is qualified and $X\cap Y=\emptyset$.
Then it holds that
$I(\mu;W_{XY})=H(\sigma_{\mu})$ and $I(\mu;W_{Y})=0$
for any $\mu\in\P_+(\S_1(\H))$
from Theorem \ref{th:QSSS-info},
and hence, we have
\begin{align}
H(\sigma_{\mu})
&=I(\mu;W_{XY}) - I(\mu;W_{Y})
\nonumber\\
&=H(W_{XY}(\sigma_{\mu})) - \EX{\mu}{H(W_{XY}(\rho))}
\nonumber\\
&\quad -H(W_Y(\sigma_{\mu})) + \EX{\mu}{H(W_{Y}(\rho))}
\nonumber\\
&\le H(W_X(\sigma_{\mu})) - \EX{\mu}{H_{\rho}(W_{X}|W_{Y})},
\label{QSSS-eval-1}
\end{align}
where the last inequality
follows from the subadditivity of the von Neumann entropy
and we have written the conditional entropy as
\begin{align}
H_{\rho}(W_{X}|W_{Y})\Def H(W_{XY}(\rho))-H(W_{Y}(\rho)).
\end{align}
Now let $Z=N\backslash(X\cup Y)$
and note that $W_N=W_{XYZ}$ is a pure state channel.
Then
it follows from Proposition \ref{th:no-cloning-deleting} in the appendix
that qualified $X\cup Y$ implies forbidden $Z$
and forbidden $Y$ implies qualified $X\cup Z$.
Hence, since $Z$ has the same property as $Y$,
$Z$ also satisfies the same inequality as \eqref{QSSS-eval-1}, i.e.,
\begin{align}
H(\sigma_{\mu})
&\le H(W_X(\sigma_{\mu})) - \EX{\mu}{H_{\rho}(W_{X}|W_{Z})}.
\label{QSSS-eval-2}
\end{align}
Since $W_{XYZ}(\rho)$ is a pure state,
we have
$H(W_{XY}(\rho))=H(W_{Z}(\rho))$ and $H(W_{XZ}(\rho))=H(W_{Y}(\rho))$.
Consequently,
it follows from \eqref{QSSS-eval-1} and \eqref{QSSS-eval-2} that
\begin{align}
H(\sigma_{\mu})
&\le H(W_X(\sigma_{\mu}))
\nonumber\\
&\quad -\frac{1}{2}\EX{\mu}{H_{\rho}(W_{X}|W_{Y})+H_{\rho}(W_{X}|W_{Z})}
\nonumber\\
&= H(W_X(\sigma_{\mu})),
\end{align}
which has been asserted.
\end{proof}
\begin{corollary}[Gottesman \cite{Gottesman}]\label{QSSS-dim-eval}
For any significant share $i\in N$ of any QSS scheme $W_N$,
we have
\begin{align}
\dim\H\le\dim\H_i.
\label{eval-dimension}
\end{align}
\end{corollary}
\begin{proof}
Let $\mu$ be the uniform distribution on $\S_1(\H)$
in Theorem \ref{th:QSSS-entropy-eval},
namely the invariant measure with respect to the special unitary group.
Then we have $\sigma_{\mu}=I/\dim\H$ and
the dimension of each share is bounded below as
\begin{align}
\log\dim\H=H(\sigma_{\mu})\le H(W_i(\sigma_{\mu}))\le \log\dim\H_i.
\end{align}
\end{proof}

\begin{remark}
\label{remark:4}
The arguments and the theorems so far are valid even
in the classical cases.
That is verified by replacing the corresponding notions
with the classical ones.
For example,
quantum operations, the Holevo information, and pure states
are replaced with
channels, the mutual information, and delta distributions, respectively.
In this case, it should be noted that
the proof of Theorem \ref{th:QSSS-entropy-eval}
is already finished in \eqref{QSSS-eval-1},
since the conditional entropy is nonnegative in the classical cases.
\end{remark}

\section{Ramp QSS Schemes}

From Corollary \ref{QSSS-dim-eval},
it is impossible
to reduce the dimension of each share than that of the original system
in perfect QSS schemes,
since any share except useless ones should be significant
in perfect QSS schemes.
On the other hand, in classical ramp SS schemes
such as $(k,L,n)$-threshold ramp SS schemes
\cite{Yamamoto,Blakley-Meadows},
the size of each share can be decreased
by taking into account the trade-off between
the security condition and the coding efficiency.
We utilize this idea in the quantum setting
to propose
\textit{$(k,L,n)$-threshold ramp QSS schemes} in the following sense.
\begin{definition}\label{def:ramp}
A QSS scheme $W_N$ is called
a $(k,L,n)$-threshold ramp QSS scheme
if the following conditions are fulfilled.
\begin{enumerate}
\item[(a)] $X\subseteq N$ is forbidden iff $|X|\le k-L$.
\item[(b)] $X\subseteq N$ is qualified iff $|X|\ge k$.
\end{enumerate}
\end{definition}
Note that the above conditions imply
\begin{enumerate}
\item[(c)] $X\subseteq N$ is intermediate iff $k-L<|X|<k$,
\end{enumerate}
and the $(k,L,n)$-threshold ramp QSS scheme reduces to
the $(k,n)$-threshold QSS scheme \cite{Cleve-Gottesman-Lo} if $L=1$.

Cleve \etal \cite{Cleve-Gottesman-Lo} showed that
the condition $n\le 2k-1$ must be satisfied 
for the $(k,n)$-threshold QSS scheme to exist.
As an extension of this condition, we have the following lemma.
\begin{lemma}\label{lem:kLn-relation}
For a $(k,L,n)$-threshold ramp QSS scheme,
it holds that $n\le 2k-L$.
Especially, we have $n=2k-L$ if it is a pure state QSS scheme.
\end{lemma}
\begin{proof}
From Definition \ref{def:ramp},
$X$ is a qualified set if $|X|= k$.
In this case,
it follows from Proposition \ref{th:no-cloning-deleting} in the appendix
that the complement $N\backslash X$ is forbidden,
which implies $|N\backslash X|=n-k\le k-L$.
In the case of a pure state QSS scheme,
we can also show that $n\ge 2k-L$ in the same way.
\end{proof}

Similarly to Theorem \ref{th:QSSS-entropy-eval}
and Corollary \ref{QSSS-dim-eval},
we can evaluate the coding efficiency of the ramp scheme as follows.
\begin{theorem}\label{th:kLn-entropy-bound}
For $(k,L,n)$-threshold ramp QSS schemes, it holds that
\begin{align}
\forall\mu\in\P_+(\S_1(\H)),\,
\oneover{L}H(\sigma_{\mu})\le\oneover{n}\sum_{i\in N}H(W_i(\sigma_{\mu})).
\label{kLn-eval-1}
\end{align}
\end{theorem}
\begin{proof}
For any set $X\subseteq N$ with the cardinality $|X|=L$,
there exists $Y\subseteq N$ such that $X\cap Y=\emptyset$
and $|Y|=k-L$. Then $X\cup Y$ is qualified while $Y$ is forbidden.
Therefore it follows from Theorem \ref{th:QSSS-entropy-eval} that
\begin{align}
H(\sigma_{\mu})
\le H(W_X(\sigma_{\mu}))
\le \sum_{i\in X} H(W_i(\sigma_{\mu})),
\label{kLn-eval-2}
\end{align}
where we used the subadditivity of the von Neumann entropy.
Finally we can show \eqref{kLn-eval-1}
by taking the arithmetic mean of \eqref{kLn-eval-2}
for all $X\subseteq N$ satisfying $|X|=L$.
\end{proof}

\begin{corollary}\label{cor:kLn-dim-bound}
For $(k,L,n)$-threshold ramp QSS schemes, we have
\begin{align}
\oneover{L}\dim\H\le\oneover{n}\sum_{i\in N}\dim\H_i.
\label{kLn-eval-3}
\end{align}
\end{corollary}
The above corollary implies that
the dimension of each share can be decreased
by the factor $1/L$ in the average sense
than that of the original system.

\section{Construction of Ramp Schemes}

In this section, we will show a method
to realize $(k,L,n)$-threshold ramp QSS schemes
which has the optimal coding efficiency in the sense of
Corollary \ref{cor:kLn-dim-bound}.
The encoding and reverse operations used here
are regarded as extensions of
Ref.~\cite{Cleve-Gottesman-Lo} to the ramp QSS scheme.

Let $\F$ be a finite field with $q\Def|\F|\ge n$,
and let $\J_j\;(j=1,\dots,L)$ and $\H_i\;(i\in N=\{1,\dots,n\})$
be isomorphic
Hilbert spaces with dimension $\dim\J_j = \dim\H_i = q$
and an orthonormal basis $\{\cket{s}\}_{s\in\F}$
indexed by $\F$.
We will construct a pure state QSS scheme $W_N$
which maps a quantum state on $\H\Def\bigotimes_{j=1}^L \J_j$
into the composite system of
shares $\H_N\Def\bigotimes_{i\in N} \H_i$.
Note that $n=2k-L$ holds from Lemma \ref{lem:kLn-relation}.
Since the pure state QSS scheme $W_N$ is represented by an isometry
$V:\H\rightarrow\H_N$ as
\begin{align}
W_N(\rho)=V\rho V^*,
\label{kLn-const-1}
\end{align}
it suffices to specify the images $V\cket{s^L}$ of the basis
\begin{align*}
\cket{s^L}=\cket{s_1}\otimes\dots\otimes\cket{s_L},\quad
s^L=(s_1,\dots,s_L)\in\F^L
\end{align*}
on $\H$.
For this purpose, we utilize
the polynomial of degree $k-1$ on $\F$
specified by coefficients $c=(c_1,\dots,c_k)\in\F^{k}$, i.e.,
\begin{align}
p_c(x)=\sum_{i=1}^k c_ix^{i-1}.
\label{kLn-const-3}
\end{align}
By providing publicly revealed constants $x_1,\dots,x_n\in\F$
which are different from each other,
define the isometry $V$ by
\begin{align}
V\cket{s^L}\Def\oneover{\sqrt{C}}\sum_{c\in D(s^L)}
\cket{p_c(x_1),\dots,p_c(x_n)},
\label{kLn-const-4}
\end{align}
where
\begin{align*}
D(s^L)\Def
\defset{(c_1,\dots,c_k)\in\F^k}{c_i=s_i\,(i=1,\dots,L)},
\end{align*}
and $C$ is a normalization constant to be specified later.
Now, in order to verify that $V$ is actually an isometry,
let us introduce the following notations
for $X=\{i_1,\dots,i_m\}\subseteq N$
\begin{align}
M_b^a(X)&\Def
\begin{pmatrix}
x_{i_1}^a & \dots & x_{i_m}^a \\
x_{i_1}^{a+1} & \dots & x_{i_m}^{a+1} \\
\vdots &  & \vdots \\
x_{i_1}^b & \dots & x_{i_m}^b
\end{pmatrix}
\quad (a<b),
\\
p_c(X)&\Def(p_c(x_{i_1}),\dots,p_c(x_{i_m})).
\label{kLn-const-6}
\end{align}
Then we have $p_c(X)=(c_1,\dots,c_k) M_{k-1}^0(X)$,
and the following lemma is useful for later discussions.
\begin{lemma}\label{lem:injective}
For each $s^L\in\F^L$,
the map $c\in D(s^L)\mapsto p_c(X)$ is injective if $|X|\ge k-L$.
Especially, it is one-to-one if $|X|=k-L$.
Similarly, the map
$c\in\F^k\mapsto p_c(X)$ is injective if $|X|\ge k$,
and it is one-to-one if $|X|=k$.
\end{lemma}
\begin{proof}
The injective property is verified by the following relation
\begin{align}
p_c(X)=(s_1,\dots,s_L,c_{L+1},\dots,c_k)
\begin{pmatrix}
M_{L-1}^{0}(X) \\
M_{k-1}^{L}(X) \\
\end{pmatrix},
\label{injective-1}
\end{align}
since $M_{k-1}^{L}(X)$ has the full column rank if $|X|\ge k-L$.
In addition, it is one-to-one if $|X|=k-L$,
since $|D(s^L)|=|\F^{k-L}|$.
In the same way, we can show
the remaining part of the lemma.
\end{proof}
From the above lemma,
we can see that
$\cket{p_c(N)}\,(c\in\F^k)$ are orthogonal to each other,
and hence,
$V\cket{s^L}\,(s^L\in\F^L)$ are also orthogonal to each other,
which ensures that $V$ is isometric.
At the same time,
the normalizing constant is determined as $C=q^{k-L}$.

Next, we will show that thus constructed QSS scheme $W_N$ is
actually a $(k,L,n)$-threshold ramp scheme.

\paragraph{Qualified sets}

In order to verify that $X$ is qualified for $|X|\ge k$,
it suffices to show that
$X=\{1,\dots,k\}$ is qualified, because of
the symmetrical way to construct $W_N$
and the monotonicity of the access structure.
The following local operations on $X$
realize the reverse operation of $W_X$.
\begin{enumerate}
\item
Perform the unitary transformation on $X$
corresponding to $p_c(X) M^0_{k-1}(X)^{-1}$,
which turns the summation in \eqref{kLn-const-4} into
\begin{align}
\sum_{c\in D(s^L)}
\cket{c_1,\dots,c_k,p_c(x_{k+1}),\dots,p_c(x_n)}.
\label{authrized-sets-1}
\end{align}
\item
Noting that $n-k=k-L$,
perform the unitary transformation on $X$ corresponding to
the linear transformation:
\begin{align}
(c_1,\dots,c_k)
\begin{pmatrix}
I & M_{L-1}^0(N\backslash X) \\
0 & M_{k-1}^{L}(N\backslash X)
\end{pmatrix}.
\label{authrized-sets-2}
\end{align}
Then \eqref{authrized-sets-1} yields
\begin{align}
\cket{s^L}\sum_{c\in D(s^L)}\cket{p_c(N\backslash X)}\cket{p_c(N\backslash X)},
\label{authrized-sets-3}
\end{align}
which can be represented by Lemma \ref{lem:injective} as
\begin{align}
\cket{s^L}\sum_{y^{k-L}\in\F^{k-L}}\cket{y^{k-L}}\cket{y^{k-L}}.
\label{authrized-sets-4}
\end{align}
\end{enumerate}
Thus, we have recovered $\cket{s^L}$ on $\{1,\dots,L\}$
from $V\cket{s^L}$ by the local operations on $X$.

\paragraph{Forbidden sets}

When $|X|\le k-L$,
$N\backslash X$ is qualified since $|N\backslash X|\ge k$.
Therefore it follows from Proposition \ref{th:no-cloning-deleting}
in the appendix that $X$ is forbidden.

\paragraph{Intermediate sets}

In the case $|X|=k-l\;(0<l<L)$, we show that
$X$ is intermediate.
For this purpose, it is sufficient to show that
for $X=\{1,\dots,k-l\}$,
$W_X$ is neither reversible nor vanishing \wrt a subset
$\S=\left\{\pure{s^L}\right\}_{s^L\in\F^L}$
included by $\S(\H)$.
Taking Theorem \ref{th:QSSS-info} into account,
let us calculate the Holevo information
\begin{align*}
I(\mu;W_X)=H(W_X(\sigma_{\mu}))
-E_{\mu}\!\left[ H\left(W_X\left(\pure{s^L}\right)\right) \right]
\end{align*}
and $H(\sigma_{\mu})$
for the uniform distribution $\mu$ on $\S$.
Then 
the von Neumann entropy is easily calculated as
$H(\sigma_{\mu})=L\log q$ since $\sigma_{\mu}=I/q^{L}$.
On the other hand,
from Lemma \ref{lem:injective} and
\begin{align*}
|N\backslash X|=n-(k-l)=(2k-L)-(k-l)> k-L,
\end{align*}
we have
\begin{align}
&W_X\left(\pure{s^L}\right)
\nonumber\\
&=\oneover{C}\sum_{c,d\in D(s^L)}
\inner{p_d(N\backslash X)}{p_c(N\backslash X)}
\cket{p_c(X)}\!\bra{p_d(X)}
\nonumber\\
&=\oneover{C}\sum_{c\in D(s^L)}
\cket{p_c(X)}\!\bra{p_c(X)}.
\label{intermediate-sets-1}
\end{align}
Lemma \ref{lem:injective} with $|X|=k-l>k-L$
also enables us to see that
$\cket{p_c(X)}$
in the summation in \eqref{intermediate-sets-1}
are orthogonal to each other,
and hence,
we have
$H\left(W_X\left(\pure{s^L}\right)\right)=(k-L)\log q$
for all $s^L\in\F^L$.
Next, letting $Y=\{1,\dots,k\}$, we have
\begin{align}
&W_Y(\sigma_{\mu})
\nonumber \\
&=\oneover{q^L}\sum_{s^L\in\F^L}W_Y\left(\pure{s^L}\right)
\nonumber \\
&=\oneover{q^L C} \!\! \sum_{s^L\in\F^L} \! \sum_{c,d\in D(s^L)}
\!\!\!\!\! \inner{p_d(N\backslash Y)}{p_c(N\backslash Y)}
\cket{p_c(Y)}\!\bra{p_d(Y)}
\nonumber \\
&=\oneover{q^k}\sum_{s^L\in\F^L}\sum_{c\in D(s^L)}
\cket{p_c(Y)}\!\bra{p_c(Y)}
\nonumber \\
&=I/q^k,
\end{align}
where the third and last equalities follow from Lemma \ref{lem:injective}.
Then we have $W_X(\sigma_{\mu})=I/q^{k-l}$
and $H(W_X(\sigma_{\mu}))=(k-l)\log q$.
Consequently, it holds that
\begin{align}
0<I(\mu;W_X)=(L-l)\log q<L\log q=H(\sigma_{\mu}).
\end{align}
Therefore it follows from Theorem \ref{th:QSSS-info} that
$W_X$ is neither reversible nor vanishing \wrt $\S$.

At last, it is confirmed that
$W_N$ actually realizes the $(k,L,n)$-threshold ramp scheme.
It is clear from the construction of $W_N$
that the coding efficiency of $W_N$ is optimal in the sense of 
Corollary \ref{cor:kLn-dim-bound}.

\section{Concluding Remarks}

In this paper,
we have revisited the lower bound on the dimension of each share
in QSS schemes given by Gottesman \cite{Gottesman},
and gave a rigorous proof for the lower bound
(Theorem \ref{th:QSSS-entropy-eval} and Corollary \ref{QSSS-dim-eval}).
The key idea of the proof was as follows.
First, we have established a fundamental relation
between the reversibility of quantum operations and
the Holevo information (Theorem \ref{th:Holevo-info}).
Then, we have treated the qualified or forbidden condition
as the reversible or vanishing condition
for the corresponding quantum operation.
These steps gave us clear insights into QSS schemes
and even into classical SS schemes (see Remark \ref{remark:4}).
For example, we can easily see from these pictures
that the qualified or forbidden condition
is independent of the probability of the source ensemble
in both classical and quantum cases (Remark \ref{remark:3}).

We have also proposed a ramp QSS scheme
called the $(k,L,n)$-threshold QSS ramp scheme
so that the dimension of each share could be decreased than
that of the original system by the sacrifice of security conditions.
Finally, we have
analyzed the coding efficiency of the $(k,L,n)$-threshold ramp scheme
and shown an optimal construction to attain the lower bound on the efficiency.

One may wonder that
a forbidden party with the forbidden set of shares
could disturb the protocol in QSS schemes
by using the property of the entanglement.
In other words, what happens if
the forbidden party would try to break the protocol
to measure their particles and to announce the outcome publicly?

The answer of the question is as follows.
Let $X\subseteq N$ be a qualified set for a QSS scheme $W_N$.
Then, $Y\Def N\backslash X$ is forbidden and
there exists a decoding operation $\R_X$ for $X$
recovering any pure state $\rho$ from $W_N(\rho)$:
\begin{align}
(\I_Y\otimes\R_X) W_N(\rho)
&= W_Y(\rho)\otimes\rho
\nonumber \\
&= \rho_{0}\otimes\rho,
\label{concluding-1}
\end{align}
where $\I_Y$ is the identity map and $\rho_{0}\Def W_Y(\rho)$.
Note that pure states have no entanglement with another systems
and that $\rho_{0}$ does not depend on $\rho$.
Then, we can see that \eqref{concluding-1} also holds
for mixed states $\rho$,
and that any malicious operation $\E_Y$ by the forbidden party $Y$
could not interfere with the qualified set of shares, i.e.,
\begin{align}
&(\I_Y\otimes\R_X) (\E_Y\otimes \I_X) W_N(\rho)
\nonumber \\
&= (\E_Y\otimes \I_X) (\I_Y\otimes\R_X) W_N(\rho)
\nonumber \\
&= \E_Y (\rho_0)\otimes\rho.
\end{align}

However, the above arguments are not valid for
intermediate sets of shares, i.e.,
a measurement on an intermediate set of shares may
affect the quantum state of another intermediate set
through the effects of the entanglement.
For this reason, it is a challenging problem
to classify intermediate sets in ramp QSS schemes.
We need to study security conditions for ramp QSS schemes
and to develop tools to
quantify the information that an intermediate set of share has.
These developments are left to further studies.

\appendix*
\section{No Cloning and No Deleting Theorem}

In this paper,
we have used a fundamental result in QSS schemes
shown by Cleve \etal \cite{Cleve-Gottesman-Lo},
that is,
if a set of shares  $X\subseteq N$ is qualified,
the complement $N\backslash X$ is necessarily forbidden,
and, in addition, the converse is also true
in pure state QSS schemes.
These properties are regarded as variants of
the no cloning theorem
\cite{Wootters-Zurek,Dieks,Yuen,Barnum-et-al,Koashi-Imoto:No-cloning}
and the no deleting theorem
\cite{Pati-Braunstein:Nature,Pati-Braunstein:quant-ph/0007121}.
Another proof of this property relevant to the no cloning theorem
is given by an information theoretical manner in Ref.~\cite{Anderson}.
In this appendix, we will review these results
in our notations for readers' convenience
following the original proof \cite{Cleve-Gottesman-Lo}
which utilizes the perfect error correcting condition
\cite{Knill-Laflamme} \cite{Bennett-et-al}.
Here we introduce a notation $\E\sim\{E_a\}_a$
by which we mean that $\E$ is a quantum operation represented by
the Kraus representation \cite{Kraus} $\E(\rho)=\sum_aE_a\rho E_a^*$.
\begin{proposition}[\cite{Knill-Laflamme,Bennett-et-al}]
\label{th:Knill-Laflamme}
Let $\C:\rho\in\S(\H)\mapsto V\rho V^*\in\S(\J)$
be a quantum operation defined by an isometry $V:\H\rightarrow\J$,
and let $\E:\S(\J)\rightarrow\S(\K)$ be a quantum operation
represented by $\E\sim\{E_a\}_a$.
Then the following conditions are equivalent.
\begin{enumerate}
\item[(a)] $\E\cdot\C$ is reversible \wrt $\S(\H)$.
\item[(b)] For each pair of indices $a$ and $b$,
there exists $C_{ab}\in\Comp$ such that
$
V^*E_a^*E_bV=C_{ab}I_{\H}
$.
\end{enumerate}
\end{proposition}

\begin{proposition}[Cleve-Gottesman-Lo \cite{Cleve-Gottesman-Lo}]
\label{th:no-cloning-deleting}
Given a quantum operation $W_{XY}:\S(\H)\rightarrow\S(\H_X\otimes\H_Y)$,
let $W_X\Def\Tr_Y\cdot W_{XY}$ and $W_Y\Def\Tr_X\cdot W_{XY}$.
If $W_{XY}$ is a pure state channel and reversible,
then the following conditions are equivalent.
\begin{enumerate}
\item[(a)] $W_X$ is reversible \wrt $\S(\H)$.
\item[(b)] $W_Y$ is vanishing \wrt $\S(\H)$.
\end{enumerate}
In the case of general quantum operations, (a) implies (b).
\end{proposition}
\begin{proof}
First, we show the equivalence
when $W_{XY}$ is a pure state channel and reversible.
In this case, $W_{XY}$ is written as $W_{XY}(\rho)=V\rho V^*$
by an isometry $V:\H\rightarrow\H_X\otimes\H_Y$.
Let $\{\cket{a}\}_a$ be an orthonormal basis on $\H_Y$.
Then it follows from Proposition \ref{th:Knill-Laflamme}
with $\E=\Tr_Y\sim\{I_X\otimes \bra{a}\}_a$ that 
(a) holds iff there exists $C_{ab}\in\Comp$ such that
\begin{align}
\forall (a,b),\,V^*(I_X\otimes\cket{a}\!\bra{b})V=C_{ab}I_{\H}.
\label{ap:no-cloning-1}
\end{align}
Moreover, \eqref{ap:no-cloning-1} is equivalent to the existence
of a linear functional $C:\L(\H_Y)\rightarrow\Comp$ such that
\begin{align}
\forall A\in\L(\H_Y),\, V^*(I_X\otimes A)V=C(A)I_{\H}.
\label{ap:no-cloning-2}
\end{align}
Now we can easily see the equivalence of
\eqref{ap:no-cloning-2} and (b) from the following equalities:
\begin{align}
\TR{W_Y(\rho)A}
&=\TR{W_{XY}(\rho)(I_X\otimes A)} \nonumber\\
&=\TR{\rho V^*(I_X\otimes A)V}.
\end{align}

In the general case,
let $W_{XY}(\rho)=\PTR{Z}{U\rho U^*}$ be the Stinespring representation
\cite{Stinespring},
where $U:\H\rightarrow\H_X\otimes\H_Y\otimes\H_Z$ is an isometry.
Then $W_{XYZ}(\rho)=U\rho U^*$ is a pure state channel and reversible.
From the above argument,
if $W_X$ is reversible \wrt $\S(\H)$, then $W_{YZ}$ is vanishing,
and hence $W_{Y}$ is also vanishing \wrt $\S(\H)$.
\end{proof}


\end{document}